# A finite element formulation of the outlet gradient boundary condition for convective-diffusive transport problems

Fabien Cornaton[1], Pierre Perrochet[1] and Hans-Jörg Diersch[2]

[1]*Centre of Hydrogeology, Institute of Geology, University of Neuchâtel, Emile-Argand 11, CH-2007 Neuchâtel, Switzerland*
[2]*WASY Institute for Water Resources Planning and System Research Ltd., Waltersdorfer Str. 105, D-12526 Berlin, Germany*

## SUMMARY

A simple finite element formulation of the outlet gradient boundary condition is presented in the general context of convective-diffusive transport processes. Basically, the method is based on an upstream evaluation of the dependent variable gradient along open boundaries. Boundary normal unit vectors and gradient operators are evaluated using covariant bases and metric tensors, which allow handling finite elements of mixed dimensions. Even though the presented method has implications for many fields where diffusion processes are involved, discussion and illustrative examples address more particularly the framework of contaminant transport in porous media, in which the outlet gradient concentration is classically, but wrongly assumed to be zero.

KEY WORDS: finite elements; open boundary problem; upstream gradient evaluation; implicit Neumann condition

## 1. INTRODUCTION

Convective-diffusive transport simulations require the prescription of specific boundary conditions. Particularly the inlet and outlet limits of a given reservoir often lead to boundary condition effects on the behaviour of the conserved property. Boundary conditions are obtained from the flux conservation principle accounting for the fact that there cannot be accumulation at the boundary [1]. The outcome of the transported dependent variable is partially linked to the kind of boundary condition that is used at inflow and outflow boundaries of the dynamic system. The latter boundary, which in many cases corresponds to an open boundary of the reservoir, is often the most delicate to handle because the convective and dispersive quantities cannot be specified a priori. In practice, a direct consequence is that outflow boundaries are often subject to the assumption that the gradient is zero [2], with the consequence that the boundary is impermeable to the normal diffusive (or dispersive) fluxes. Usually, the assignment of such condition at outlet derives from technical advantages of resolution or intuitive choices, rather than from physical considerations and field observations. However, for finite reservoirs one must evaluate the influence of the exit boundary on the upstream behaviour of the transported property.

Various transport column experiments in porous media clearly showed that the physical meaning of convective flux permeable and dispersive flux impermeable limits is not always obvious [3-8]. More formally speaking, the representation of boundaries in mathematical continua put forward the presence of a finite and very small transition zone (Figure 1) within which the medium properties vary continuously, ensuring macroscopic mass balance and a self-consistent definition of the transported property [3], provided the fact that a total flux is prescribed. Integration of the mass conservation equation over the finite transition zone leads to continuity of the total flux [3]. As discussed by Nauman and Buffham [9], the formulation of boundaries permeable to both the convective and the dispersive parts of the total flux permits upgradient solute movement by dispersion. Parker [6] and Novakowski [7,8] provided experimental data supporting the meaning of a total flux formulation at outflow boundaries.

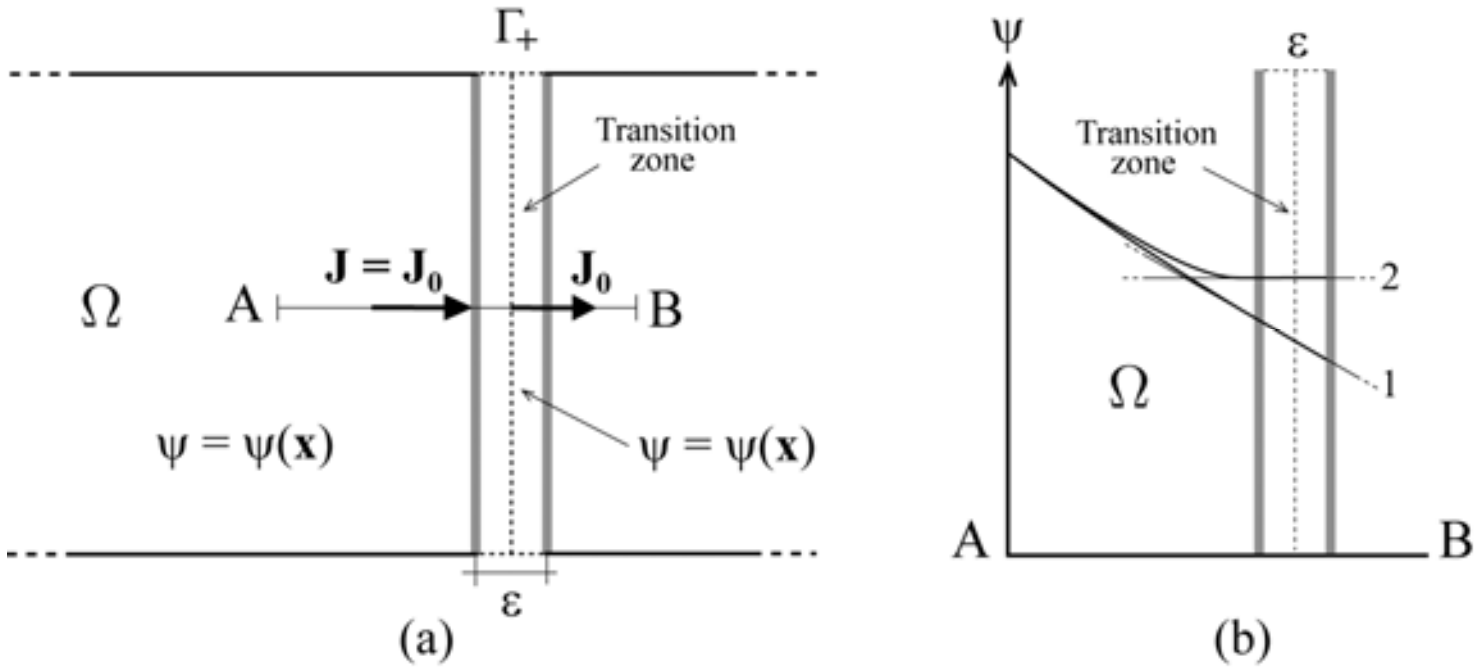

Figure 1. Finite transition zone representation of the out flowing open boundary $\Gamma_+$ of a medium $\Omega$: (a) Continuity of the mass flux **J** within the transition zone of infinitesimal size $\varepsilon$, inside which the property $\psi = \psi(\mathbf{x})$ may vary continuously; (b) Profile showing the behaviour of the property when it varies continuously (1) and when gradient is forced to zero (2).

The conceptualization of a zero gradient or homogeneous Neumann condition at outlet boundaries, which is also called the 'natural' or also the Danckwerts condition [10], comes together with the assumption that the existence of a boundary layer at the outlet may not be realistic, and that the effluent boundary should not influence the property within the medium. However, the macroscopic treatment of boundaries implies that continuity of the transported property at a microscopic level has poor relevance when volume-averaged equations are used. The irregularity of the medium structure at a microscopic level may alter the validity of the Danckwerts condition, which assumes that the volume-averaged property is equivalent to the flux-averaged property. When dispersion processes are included into transport phenomena, no clear physical evidence can support the idea of a zero concentration gradient at the interface between the considered medium and the surroundings. Moreover, the diffusive component of the flux at the outlet cannot be dropped without losing the generality of the transport equations. As example consider the common situations encountered in sub-surface hydrology, of groundwater volumes flowing out in reservoirs of free water like lakes. If the effluent concentration is perfectly mixed we might accept the possibility of no gradient within the transition zone. Such a situation may be encountered at ponds or mass accumulation areas due to extensive evaporation. But as long as smooth variability of parameters and conserved property can occur, a discontinuity may generally exist and permit non-zero gradient. Moreover, a property discontinuity at the outlet boundary has sense when diffusivity in the medium is important. The high diffusive effects, which are known to induce upstream mixing, could not be simulated if the gradient is forced to be zero.

As will be shown in the following, the hypothetic behaviour of a property in the neighbourhood of outlet boundaries can be a priori estimated by an upstream formulation of the normal gradients, allowing the classical assumption of a zero gradient to be removed. In Section 2 we set the considered mathematical models and boundary conditions. In Section 3 we derive finite element formulations for the outlet normal diffusive flux vector along outflow open boundaries. Finally, a theoretical illustration of the effects of the outlet boundary condition on the distribution of a solute concentration is given in Section 4. The proposed finite element formulation of open boundary problem is expected to be applicable to a wide range of engineering modelling problems where gradient-type boundary conditions require special attention.

## 2. BASIC EQUATIONS

Let us consider a bounded domain $\Omega \subset \mathbb{R}^d$, $d = 1$, 2 or 3. The domain boundary $\partial\Omega = \Gamma_1 \cap \Gamma_2 \cap \Gamma_3$ is decomposed into portions with essential (Dirichlet) and natural (Neumann) boundary conditions on $\Gamma_1$ and $\Gamma_2$, respectively, with $\Gamma_3 \neq \varnothing$ being the open boundary part of $\partial\Omega$ (see Figure 2). The classical convection-diffusion-reaction equation for a scalar space-time property $\psi = \psi(\mathbf{x}, t)$ can be expressed in three dimensions by

$$\frac{\partial \psi}{\partial t} + \mathbf{v} \cdot \nabla\psi - \nabla \cdot \mathbf{D}\nabla\psi + \alpha\psi = f \qquad \text{in} \quad \Omega, \quad t \in (0, T] \tag{1}$$

where $t$ is time in the time interval of interest $(0, T]$, and where $\nabla$ denotes the gradient, $\mathbf{v}$ is the advection velocity, $\mathbf{D}$ is a symmetric positive dispersion-diffusion tensor, $\alpha$ is a reaction function, and $f$ stands for a source/sink term. Initial conditions and standard boundary conditions yielding solutions of (1) can be formalized by the following:

$$\psi = \psi_0 \qquad \text{in} \quad \Omega, \quad t = 0 \tag{2}$$

$$\psi = \psi_1 \qquad \text{on} \quad \Gamma_1 \tag{3}$$

$$\mathbf{D}\nabla\psi \cdot \mathbf{n} + \gamma\psi = g \qquad \text{on} \quad \Gamma_2 \tag{4}$$

Equation (3) is the classical Dirichlet boundary condition, with $\psi_1$ being a prescribed value of the unknown function $\psi$. Equation (4) has the form of a Robin boundary condition from which more specific Neumann and Cauchy type conditions can be derived. In Equation (4), **n** is a normal positive outward unit vector, $\gamma$ and $g$ are functions on $\Gamma_2$. The unknown term $\mathbf{D}\nabla\psi \cdot \mathbf{n}$ at open outflow boundaries is explicitly formulated in the following using a finite element analysis.

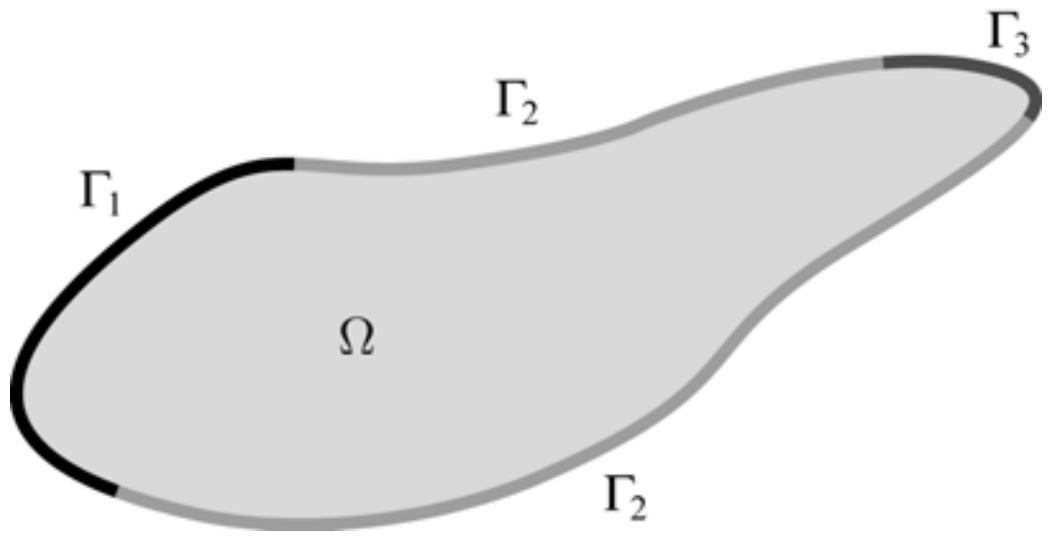

Figure 2. Schematic illustration of the considered single domain $\Omega$ and its boundary parts. $\Gamma_1$ and $\Gamma_2$ represent the boundary portions of $\partial\Omega$ where Dirichlet and Neumann conditions, respectively, are prescribed, and $\Gamma_3$ represents the open boundary portion of $\partial\Omega$.

## 3. FINITE ELEMENT FORMULATIONS

### *3.1. Weak form of the transport equation*

In preparation to the formulation of normal gradients at open outflow boundaries, we proceed to formulate the transport equation in weak form. To obtain a weak form of the boundary value problem defined by the differential equation (1), with the initial conditions (2) and the boundary conditions (3) – (4), we consider the two following spaces of weighting functions $\eta$ and solution functions $\psi$:

$$V = \left\{ \eta \in H^1(\Omega) \,\middle|\, \eta = 0 \quad \text{on } \Gamma_1 \right\}$$

$$S = \left\{ \psi \in H^1(\Omega) \,\middle|\, \psi = \psi_1 \text{ on } \Gamma_1 \right\}$$

where $H^1(\Omega)$ is the usual Sobolev space of functions which are square integrable, and which have square integrable first derivatives. The weak form is obtained by finding $\psi \in S$ such that for all $\eta \in V$

$$\left( \frac{\partial\psi}{\partial t}, \eta \right) + a(\psi, \eta) = \varphi(\eta) \qquad \text{in} \quad \Omega \tag{5}$$

with

$$\left( \frac{\partial\psi}{\partial t}, \eta \right) = \int_\Omega \frac{\partial\psi}{\partial t} \eta \, d\Omega \tag{6}$$

$$a(\psi, \eta) = \int_\Omega [(\mathbf{v} \cdot \nabla\psi)\eta + \mathbf{D}\nabla\psi\nabla\eta] \, d\Omega - \int_{\Gamma_3} (\mathbf{D}\nabla\psi \cdot \mathbf{n})\eta \, d\Gamma + \int_\Omega \alpha\psi\eta \, d\Omega \tag{7}$$

$$\varphi(\eta) = \int_\Omega f\,\eta \, d\Omega + \int_{\Gamma_2} (g - \gamma\psi)\eta \, d\Gamma \tag{8}$$

where use has been made of the Green's theorem for integrating by parts the diffusive term. A standard Galerkin finite element formulation is used to obtain a discrete problem of the weak form (5). The Dirichlet constraint in Equation (3) must be imposed on $\Gamma_1$ to Equation (8). The second term in Equation (7) exhibits the normal diffusive term $-\mathbf{D}\nabla\psi \cdot \mathbf{n}$ on the open boundary $\Gamma_3$. In the following, we propose an evaluation method of the diffusive term on the open boundary $\Gamma_3$. Usually, the outflow diffusive flux is not known necessarily, and one uses the homogeneous condition $-\mathbf{D}\nabla\psi \cdot \mathbf{n} = 0$ as a way of truncating the physical domain at a substantial distance from the zone of interest.

### *3.2. Normal gradient evaluation along open boundaries*

We evaluate the normal projection of the diffusive flux $\mathbf{J}_D = -\mathbf{D}\nabla\psi$ approaching the open boundary $\Gamma_3$ from the interior of the considered domain $\Omega$. The projection of the total diffusive flow leaving a boundary element $e$ can be implicitly formulated as a function of the $m$ nodal unknowns $\boldsymbol{\psi} = [\psi_1\ \psi_2 \ldots \psi_m]^T$ of the element:

$$Q_D^e = \sum_{1\le n\le n_b} \int_{\Gamma_e} N_n \mathbf{J}_D \cdot \mathbf{n}\, d\Gamma = \sum_{1\le n\le n_b} \int_{\Gamma_e} N_n \mathbf{D}\mathbf{B}^T\boldsymbol{\psi} \cdot \mathbf{n}\, d\Gamma = \sum_{1\le n\le n_b} \mathbf{q}_n \boldsymbol{\psi} \tag{9}$$

with $\mathbf{B}^T = \nabla\mathbf{N}^T$ denoting the transpose of the gradient matrix, and $n_b$ being the number of nodes of the considered element boundary. The nodal shape functions $\mathbf{N} = [N_1\ N_2 \ldots N_m]^T$ are defined with respect to the local coordinates $s^k$ ($s^1 = s, s^2 = t, \ldots$), as well as their partial derivatives $\nabla\mathbf{N}$ and the differential surface (3-D) or length (2-D) $d\Gamma$. Equation (9) evaluates the contribution of element $e$ to the control volumes of the $n_b$ nodes on the element boundary $\Gamma_e$. The projected diffusive out flux at a node $n$ is the product $\mathbf{q}_n\boldsymbol{\psi}$ of the vector $\mathbf{q}_n = [q_1\ q_2 \ldots q_m]^T$ with the vector of the element $m$ nodal unknowns $\boldsymbol{\psi}$. The vector $\mathbf{q}_n$ stores the $m$ components of the projected gradient vector $-\mathbf{D}\nabla()$ contributing to the out flux at node $n$. The components of $\mathbf{q}_n$ are only functions of the global coordinates of the $m$ nodes, and of the medium diffusive property $\mathbf{D}$.

To solve Equation (9) one must identify the boundary normal vector $\mathbf{n}$, which is by definition orthogonal to the element edge or face $\Gamma_e$. The gradient matrix $\mathbf{B}^T$ and the diffusion tensor $\mathbf{D}$ also require evaluation at the same points. To do so, it is convenient to operate in the local coordinate space. To describe these operations we make use of the geometrical framework of covariant bases and contravariant metric tensors, as described in books like Ciarlet [11], or more recently in the finite hyper-elements framework proposed by Perrochet [12]. This gradient generalized mapping method, which allows handling finite elements of mixed dimensions, is described in detail in Perrochet [12], and has been recently discussed by Juanes *et al.* [13]. We recall here the main results that are used in the present work, and restrict the method to the three dimensions of space only. The gradient operator $\nabla$ in a curvilinear element is expressed by the tensor product of the element covariant base $\mathbf{a}$ and the contravariant components of the gradient $\ddot{\nabla}$ in the curvilinear system. The contravariant components of $\ddot{\nabla}$ are themselves obtained by transformation of its covariant components $\nabla^*$ in the element orthogonal local system. From the covariant base

$$\mathbf{a} = \mathbf{x}^T\nabla^*\mathbf{N} = [a_{ik}] \quad , \quad a_{ik} = \frac{\partial x^i}{\partial s^k} = \sum_m \frac{\partial N_m}{\partial s^k} x_m^i \quad , \quad k_{\max} \le i_{\max} \tag{10}$$

which is the differentiation of the global coordinates of the $m$ nodes $\mathbf{x} = [x_{mi}] = [x^i{}_m,\ i = 1, 2, 3]$ with respect to the $k$ local coordinates $s^k$, and from the covariant metric tensor $\mathbf{h} = \mathbf{a}^T\mathbf{a}$, the gradient matrix and the differential volume $d\Omega$ are fully defined according to

$$\mathbf{B}^T = \nabla\mathbf{N}^T = \mathbf{a}\,\mathbf{g}\nabla^*\mathbf{N}^T = \mathbf{a}(\mathbf{a}^T\mathbf{a})^{-1}\nabla^*\mathbf{N}^T \tag{11}$$

$$d\Omega = \sqrt{\det\mathbf{h}}\, d\Omega^* \tag{12}$$

with the superscript $^*$ indicating operations in the local domain, and $\mathbf{g} = \mathbf{h}^{-1}$ being the contravariant metric tensor. The gradient operator in the curvilinear system is given by $\ddot{\nabla} = \mathbf{g}\nabla^*$.

Considering the element boundary $\Gamma_e$, its normal unit vector $\mathbf{n}$ and differential $d\Gamma_e$ in the global space can be expressed by mapping the corresponding normal unit vector and differential in the local space:

$$\mathbf{n} = \frac{\mathbf{a}\,\mathbf{g}\,\mathbf{n}_\xi^*}{\left\|\mathbf{a}\,\mathbf{g}\,\mathbf{n}_\xi^*\right\|} \tag{13}$$

$$d\Gamma_e = \sqrt{\det \mathbf{h}}\, d\Gamma^*_\xi = \sqrt{\det \mathbf{h}}\, ds^i ds^j \tag{14}$$

with $\mathbf{n}^*_\xi$ denoting the outward unit vector in the local space, oriented in the $\xi^{th}$ local coordinate direction, and $d\Gamma^*_\xi$ being the differential of the line or surface element in the local space, for which $s^\xi$ is the orthogonal local coordinate. Using Equations (11), (13) and (14) we can formulate the normal dispersive flux $\mathbf{q}_n$ of Equation (9) for node $n$ by the following:

$$\mathbf{q}_n = \int_{\Gamma_e} N_n \mathbf{D}\mathbf{B}^T \cdot \mathbf{n}\, d\Gamma_e = \int_{\Gamma_e} N_n [\, \mathbf{Dag}\, \nabla^* \mathbf{N}^T ] \cdot \mathbf{n} \sqrt{\det \mathbf{h}}\, d\Gamma^*_\xi \tag{15}$$

In Equation (15), integration is performed on the fictive one-dimension reduced element that belongs to the outlet boundary portion (see Figure 3). Therefore, each quantity is evaluated by fixing the $\xi^{th}$ local coordinate. Once $\mathbf{q}_n$ is known for node $n$, it can easily be handled during the assembling procedure like a reaction term, with the result that the coefficients in line $n$ of the global stiffness matrix $\mathbf{A} = [A_{i,j}]$ are updated according to the $m$ coefficients $q_j$ of $\mathbf{q}_n$, $A_{i=n,j} = A_{i=n,j} - q_j$, with $j = 1,\ldots, m$.

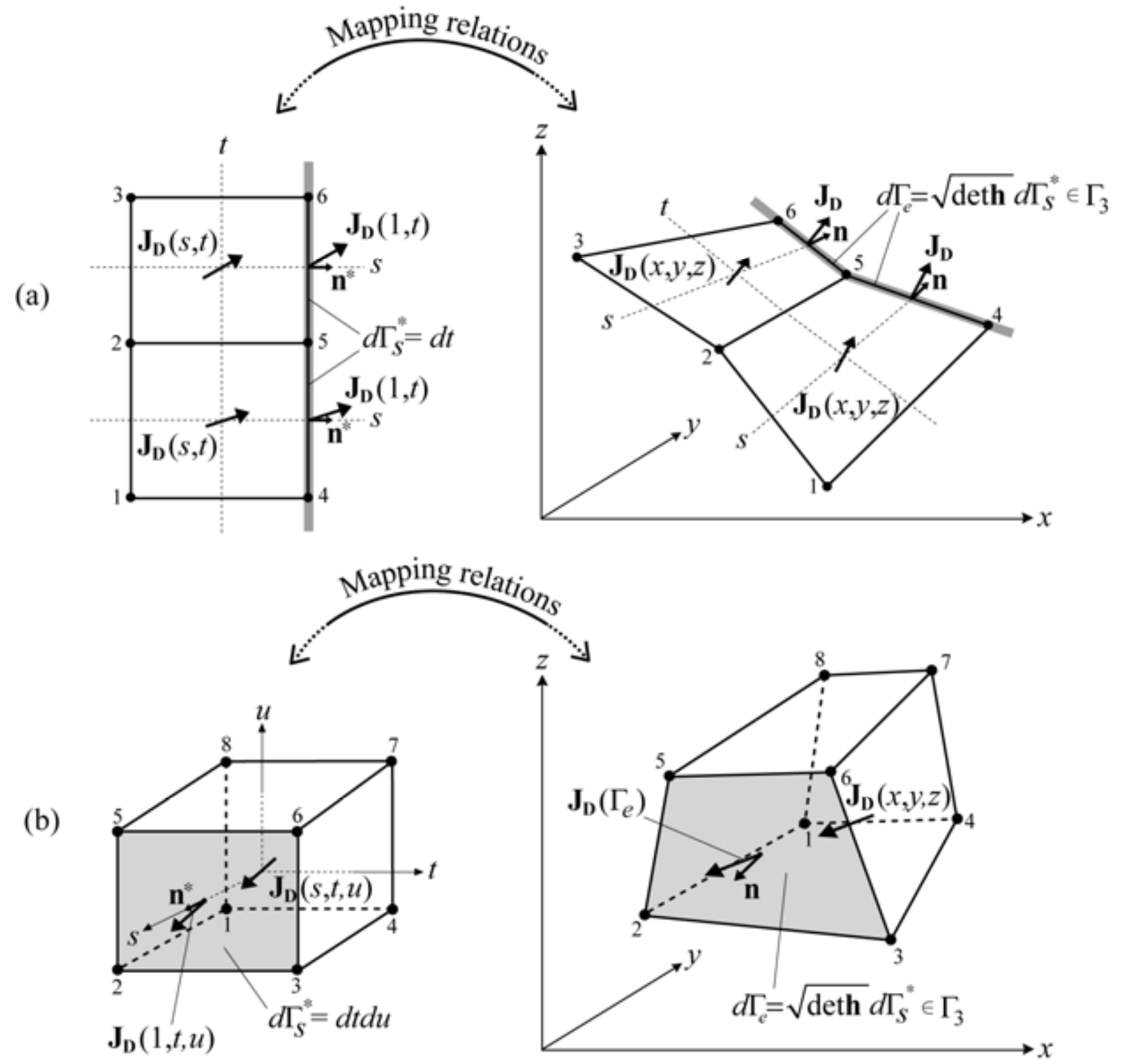

Figure 3. Examples of finite elements in the local space $(s, t, u) = (s^1, s^2, s^3)$ and the global space $(x, y, z) = (x^1, x^2, x^3)$: (a) 2-D $Q_1$ elements; (b) 3-D $Q_1$ element.

Equation (15) performs an upstream evaluation of the gradient projection along open boundaries. Therefore it comes with the implicit assumption of smooth spatial evolution of the gradients between the interior and the outside of the element. This assumption is coherent with the continuity of total flux at macroscopic level and continuity of the property at microscopic level at the boundary layer, but it also implies that beyond the medium boundary, in its neighbourhood, the gradient should be a continuation of the gradient inside the medium. The accuracy of the gradient evaluation will of course be dependent on the refinement at the boundary. Since no specific value is prescribed for the dispersive out fluxes, and since their formulation is made implicitly, we may refer this formulation of flux condition on open boundaries to as *implicit Neumann* condition. A simple example with an explicit resolution of Equation (15) is given in Appendix A, for the specific case of bilinear $Q_1$ elements.

## 3. APPLICATION

To illustrate the proposed method of implementation of realistic gradient boundary condition on open boundaries, we simulate the transport of a conservative tracer resulting from a solute and water injection. The flow and convection-dispersion equations are solved within a 2-D $(x, y)$ horizontal domain, $\Omega = [0, 500]\times[0, 500]$ (see Figure 4a), discretized using homogeneous bilinear $Q_1$ elements of size $\Delta x = \Delta y = 2$m. Flow is divergence-free, $\nabla \cdot \mathbf{v} = 0$. The boundary portion $\Gamma_0$ is a no-flow boundary ($\Gamma_2 = \Gamma_0$). Dirichlet type conditions are prescribed on the inlet boundaries $\Gamma_-$ and $\Gamma_w$ ($\Gamma_1 = \Gamma_- \cap \Gamma_w$), for both the flow and transport equations. A nil concentration is fixed on $\Gamma_-$ and a unit concentration is fixed on $\Gamma_w$. The outlet boundary $\Gamma_+$ is the open boundary, $\Gamma_3 = \Gamma_+$. A hydraulic head difference of 5m is maintained constant between $\Gamma_-$ ($H = 5$m) and $\Gamma_+$ ($H = 0$m), and $H = 3$m on $\Gamma_w$. The hydraulic head field is given in Figure 4b, and a representation of the steady-state flow velocity field is given in Figure 4c, as well as a set of path lines in Figure 4d. The velocity norm varies between 0.8 and 1m/day in undisturbed regions. The solutions of the convection-dispersion equation are tested for two situations: (*i*) with the classical homogeneous Neumann condition at outlet; (*ii*) with the *implicit Neumann* condition at outlet. The corresponding boundary value problems can be formalized by:

$$\frac{\partial C}{\partial t} + \mathbf{v}\cdot\nabla C - \nabla\cdot\mathbf{D}\nabla C = 0 \qquad \text{in} \quad \Omega \tag{16}$$

$$C(\mathbf{x}, t=0) = 0 \qquad \forall \mathbf{x} \in \Omega$$

$$C = 1 \qquad \text{on} \quad \Gamma_w$$

$$C = 0 \qquad \text{on} \quad \Gamma_-$$

and

$$(i) \qquad \mathbf{D}\nabla C \cdot \mathbf{n} = 0 \qquad \text{on} \quad \Gamma_+ \cap \Gamma_0$$

or

$$(ii) \qquad \mathbf{D}\nabla C \cdot \mathbf{n} = 0 \qquad \text{on} \quad \Gamma_0$$

$$\textit{implicit Neumann} \qquad \text{on} \quad \Gamma_+$$

In Equation (16), $C$ is relative concentration [–], $\mathbf{v}$ is the velocity field vector [m/s], and $\mathbf{D}$ [m$^2$/s] is the time-independent macro-dispersion tensor

$$\mathbf{D} = (\alpha_\mathrm{L} - \alpha_\mathrm{T})\frac{\mathbf{v}\otimes\mathbf{v}}{\|\mathbf{v}\|} + \alpha_\mathrm{T}\|\mathbf{v}\|\mathbf{I} + D_\mathrm{m}\mathbf{I} \tag{17}$$

where $\alpha_\mathrm{L}$ [m] and $\alpha_\mathrm{T}$ [m] are the longitudinal and transversal coefficients of dispersivity, respectively, $D_\mathrm{m}$ is the coefficient of molecular diffusion [m$^2$/s], and $\mathbf{I}$ is the identity matrix. The time discretization for the simulations makes use of a standard Crank–Nicholson finite-difference scheme with a constant time-step $\Delta t = 1$day.

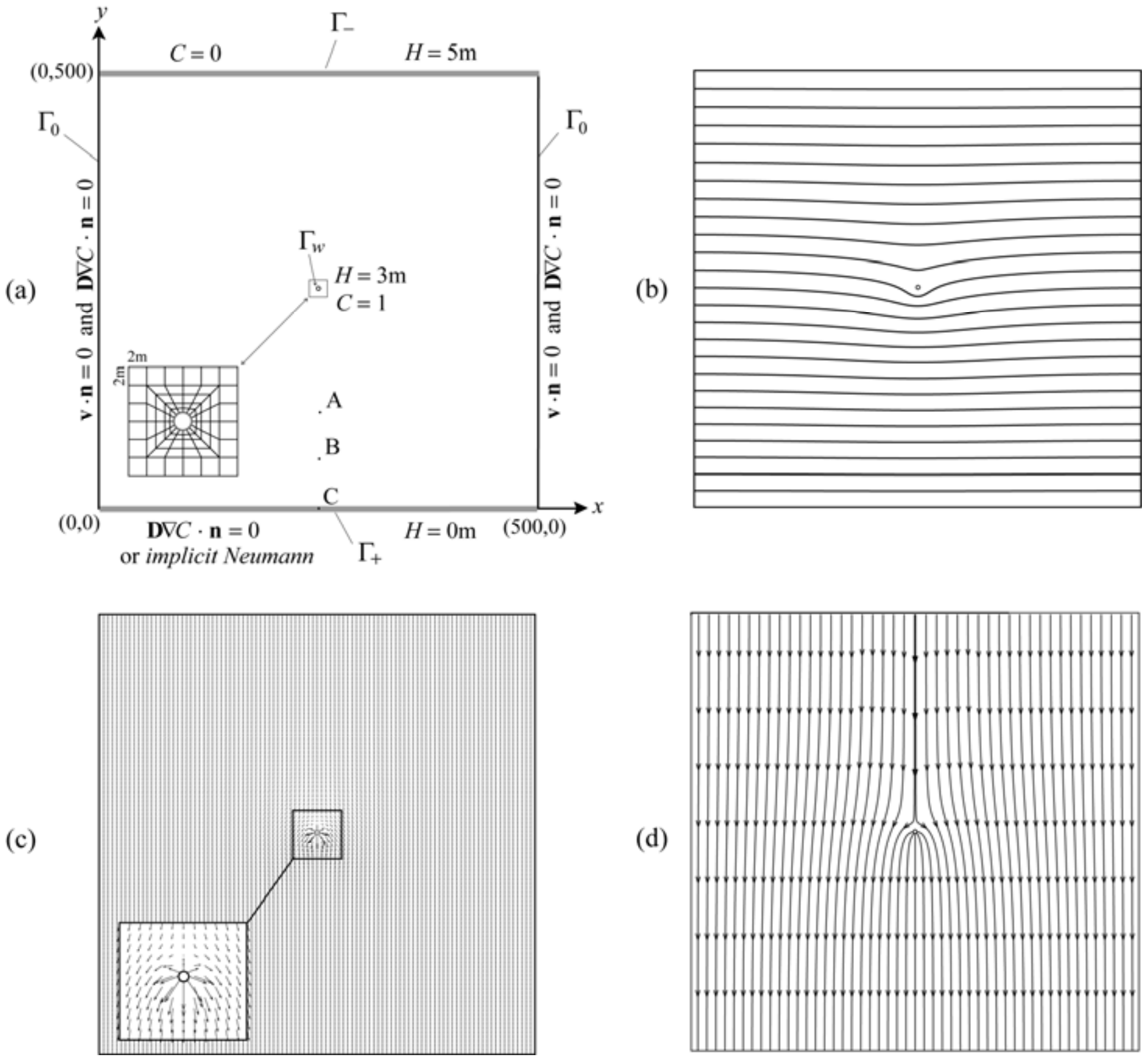

Figure 4. Definition of the 2-D $(x, y)$ flow and chemical transport problem: (a) Domain definition with its specific boundary portions and associated boundary conditions for flow and transport. A zoom of the finite element mesh around $\Gamma_w$ is given. The points A = (250, 100), B = (250, 50) and C = (250, 0) are three observation points; (b) Hydraulic heads distribution with 0.2m increments; (c) Pore velocity field; (d) Path line representation of the velocity field.

The distribution of concentration in space at time $t = 300$days is given in Figure 5. Four levels of dispersion are tested, by keeping the ratio $\alpha_L/\alpha_T$ equal to 10, and with a uniform coefficient of molecular diffusion $D_m = 2.3\times10^{-9}$ m$^2$/s (~ effective self-diffusion of water). The effect of the classical homogeneous Neumann boundary condition at outlet is clearly apparent; the iso-contours of concentration approaching $\Gamma_3 = \Gamma_+$ are forced to become perpendicular to the boundary. Without this constraint (with *implicit Neumann* on $\Gamma_3$), the same iso-contours naturally intercept the outlet boundary. The effect of the homogeneous Neumann condition on $\Gamma_3$ on the calculated concentration distribution becomes important when dispersion is increased. A direct consequence of this condition is an artificial mass accumulation at the outlet surroundings.

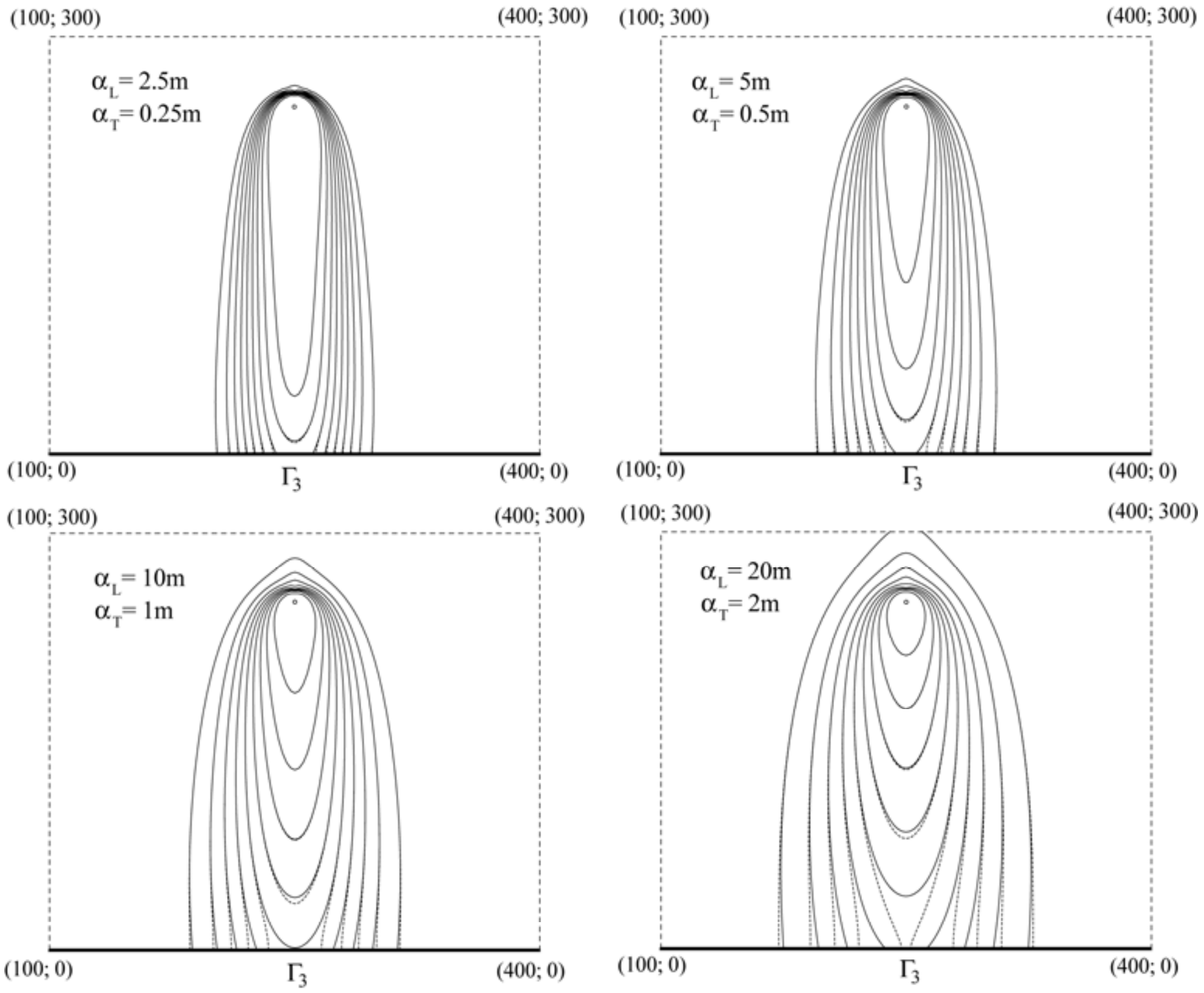

Figure 5. Compared transport solutions at time $t = 300$days, for four levels of dispersion (with $D_m = 2.3\square 10^{-9}$ $m^2$/s). The solid lines are the solution with homogeneous Neumann condition at outlet, and the dashed lines are the solution with *implicit Neumann* condition at outlet. Iso-contours of concentration from 0.1 to 0.9 with 0.1 of increment.

Observed breakthrough curves are given in Figure 6, at the three observation points A, B and C (see Figure 4a for their location). They show the effect of the homogeneous Neumann condition on the temporal evolution of concentration, at the outlet boundary and upstream inside the medium. When dispersion increases, the homogeneous Neumann condition modifies significantly the behaviour of concentration at the outlet boundary and within the flow domain. For the maximum tested dispersion case ($\alpha_L = 20$m, $\alpha_T = 0.2$m), the effect of this condition is effective until 100m upstream the outlet boundary (point A). The un-natural mass accumulation at the outlet surroundings induced by the homogeneous Neumann condition on $\Gamma_+$, is well-visible when one follows the evolution of points A, B and C against increasing dispersion. For the lowest dispersion case (Figure 6, up), the influence of the homogeneous Neumann condition does not reach point B nor A. When dispersion is increased (Figure 6, middle and down), the time-series recorded at points B and C tend to become similar.

Sensitivity analysis showed that important changes in the coefficient of molecular diffusion (from $10^{-9}$ to $10^{-6}$ $m^2$/s) do not change the concentration distribution significantly. The effect on transport solutions of the homogeneous Neumann condition prescribed on open boundaries can thus be expected to be important in systems with significant mechanical dispersion.

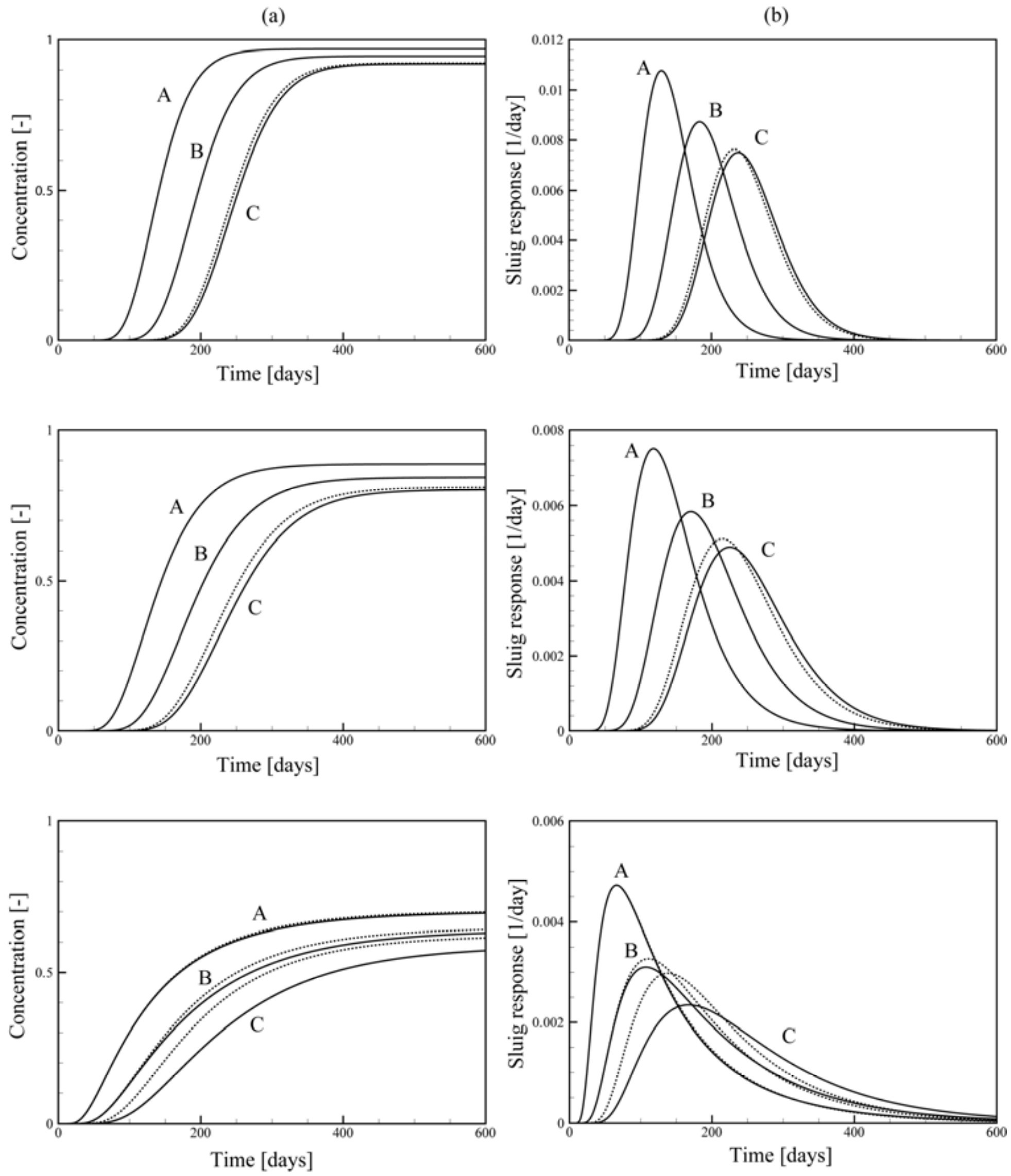

Figure 6. Compared transport solutions at the three observation points A, B and C (see Figure 4a for their location), for three levels of dispersion. From up to down: $\alpha_L = 2.5$m, $\alpha_L = 10$m, $\alpha_L = 20$m. The ratio $\alpha_L/\alpha_T$ is fixed to 10. The solid lines are the solution with homogeneous Neumann condition at outlet, and the dashed lines are the solution with *implicit Neumann* condition at outlet. (a) Observed breakthrough curves; (b) Derivatives of the observed time series (slug problem equivalent solutions).

## 4. FINAL REMARKS

The proposed outlet gradient estimation method presents the advantage that it can directly be incorporated within an element matrix integration procedure, as it requires no more than the evaluation of the classical functions. The method is straightforward for one-, two- and three-dimensional medium configurations when using covariant basis and contravariant metric tensors, which allow working simultaneously with elements of mixed dimensions. Extension of the presented finite element formulation to finite volume or finite difference schemes can also be considered.

This simple computational procedure can be useful to solve a large series of convective-dispersive problems, by treating outflow limits without taking the risk of making physically inconsistent hypothesis on the property behaviour at outlet, like the classical assumption of convective flux permeable and dispersive flux impermeable boundaries. For many transport settings occurring in finite reservoirs, like e.g. heat, mass, or residence time transport processes, the classical arbitrary homogeneous Neumann condition at outlet boundaries is therefore not needed anymore.

## AKNOWLEDGEMENTS

The authors would like to acknowledge the Swiss Research National Fund for financially supporting this research under Grant no. 2100-064927, and two anonymous reviewers who subsequently improved the quality of this paper.

## APPENDIX

An explicit resolution of Equation (15) is detailed below, using a plane domain in the Cartesian coordinate system $(x, y, z)$. The system can be assimilated to a fracture in the three dimensional space, and is discretized using three bilinear $Q_1$ elements. In Figure A1, the geometry, the node numbering and the boundary conditions are indicated. The elements node spacing is fixed to $\Delta x$ in the $x$ direction and $\Delta y$ in the $y$ direction. We want to solve the following equation:

$$-\mathbf{v} \cdot \nabla \psi + \nabla \cdot \mathbf{D} \nabla \psi + 1 = 0 \tag{A1}$$

where $\psi = \psi(x, y, z)$. Equation (A1) is a steady-state form of Eq. (1), with $\alpha = 0$ and $f = 1$. The velocity $\mathbf{v} = [v_x \ \ v_y \ \ v_z]$ is assumed to be uniform in the fracture plane, with same intensity in the $x$ and $z$ directions ($v_x = v_z = v$ and $v_y = 0$), with the result that the system behaviour will be one-dimensional in each element. Dispersion is assumed to be only controlled by molecular diffusion $\mathbf{D} = D_\mathrm{m}\,\mathbf{I}$, $\mathbf{I}$ being the identity matrix. At the two upstream nodes a Dirichlet condition is prescribed with a constant value of 0 ($\Gamma_1$ = edge 1–2 in Figure A1), $\psi(0, y, 0) = 0$. The boundary portion $\Gamma_2$ is a no-flow boundary, $\partial\psi/\partial\mathbf{n} = 0$ and $\mathbf{v} \cdot \mathbf{n} = 0$.

This problem is known as the average residence time transport, for which the exact 1-D solution $\psi(r) = r/v$ in the curvilinear coordinate $r$ of the element (in the velocity direction) is dispersion independent. The above differential equation is discretized and solved for the cases $\partial\psi/\partial\mathbf{n} = 0$ and $\partial\psi/\partial\mathbf{n} \neq 0$ at the outlet limit (edge 7–8 in Figure A1), which is the open boundary $\Gamma_3$ of the system.

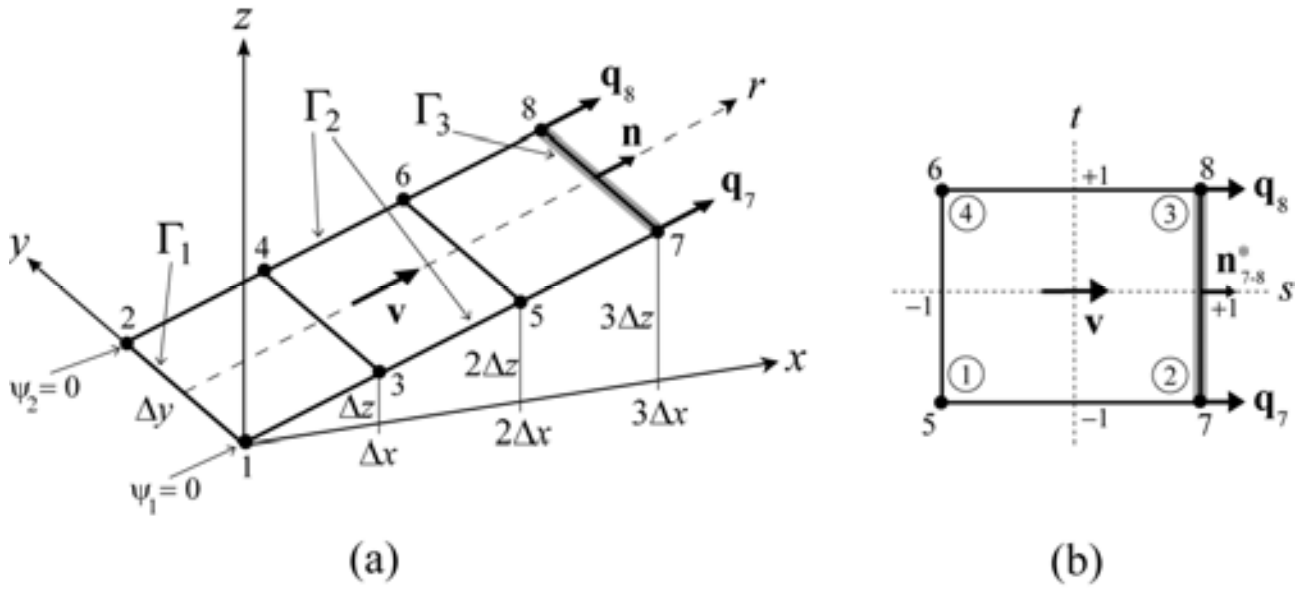

Figure A1. Discretized 2-D domain in the global Cartesian space coordinates $(x, y, z)$: (a) Geometry made of three bilinear $Q_1$ elements in the global space with prescribed Dirichlet boundary conditions at nodes 1 and 2, the numbers 1 to 8 relating the node numbering; (b) Bilinear $Q_1$ element in the local space $(s, t)$ with the numbers 1 to 4 into circles corresponding to the node numbering in the local space.

According to Equation (10), the element covariant base is

$$\mathbf{a} = \begin{bmatrix} \mathbf{a}_1 & \mathbf{a}_2 \end{bmatrix} = \begin{bmatrix} \dfrac{\partial x}{\partial s} & \dfrac{\partial x}{\partial t} \\ \dfrac{\partial y}{\partial s} & \dfrac{\partial y}{\partial t} \\ \dfrac{\partial z}{\partial s} & \dfrac{\partial z}{\partial t} \end{bmatrix} = \begin{bmatrix} x_1 & x_2 & x_3 & x_4 \\ y_1 & y_2 & y_3 & y_4 \\ z_1 & z_2 & z_3 & z_4 \end{bmatrix} \begin{bmatrix} \dfrac{\partial N_1}{\partial s} & \dfrac{\partial N_1}{\partial t} \\ \dfrac{\partial N_2}{\partial s} & \dfrac{\partial N_2}{\partial t} \\ \dfrac{\partial N_3}{\partial s} & \dfrac{\partial N_3}{\partial t} \\ \dfrac{\partial N_4}{\partial s} & \dfrac{\partial N_4}{\partial t} \end{bmatrix} = \frac{1}{2} \begin{bmatrix} \Delta x & 0 \\ 0 & \Delta y \\ \Delta z & 0 \end{bmatrix} \tag{A2}$$

for the three elements of Figure A1. The covariant metric tensor $\mathbf{h}$ reads

$$\mathbf{h} = \mathbf{a}^{\mathrm{T}}\mathbf{a} = \frac{1}{4}\begin{bmatrix} \Delta x^2 + \Delta z^2 & 0 \\ 0 & \Delta y^2 \end{bmatrix} \quad , \quad \sqrt{\det \mathbf{h}} = \frac{\Delta y}{4}\sqrt{\Delta x^2 + \Delta z^2} \tag{A3}$$

from which the contravariant metric tensor **g** becomes

$$\mathbf{g} = \mathbf{h}^{-1} = \begin{bmatrix} \dfrac{4}{\Delta x^2 + \Delta z^2} & 0 \\ 0 & \dfrac{4}{\Delta y^2} \end{bmatrix} \tag{A4}$$

The normal dispersive fluxes at the two outlet nodes are calculated by straightforward application of Equation (15) in the local coordinates ($s$, $t$):

$$\mathbf{q}_{n=7,8} = \frac{\Delta y}{2} \lim_{s=+1} \int_{-1}^{+1} N_n \mathbf{D}\mathbf{B}^{\mathrm{T}} \cdot \mathbf{n}_{7-8}\, dt \tag{A5}$$

with $\mathbf{n}_{7\text{-}8}$ the unit vector normal to edge 7–8. Enforcing Equation (13), the normal vector of edge 7–8 is

$$\mathbf{n}_{7-8} = \frac{\mathbf{a}\mathbf{g}\,\mathbf{n}_s^*}{\left\|\mathbf{a}\mathbf{g}\,\mathbf{n}_s^*\right\|} = \begin{bmatrix} \dfrac{\Delta x}{\sqrt{\Delta x^2 + \Delta z^2}} & 0 & \dfrac{\Delta z}{\sqrt{\Delta x^2 + \Delta z^2}} \end{bmatrix}^{\mathrm{T}} \tag{A6}$$

Following Equation (11) the product $\mathbf{DB}^{\mathrm{T}}$ results in

$$\mathbf{DB}^{\mathrm{T}} = \mathbf{Dag}\nabla^* \mathbf{N}^{\mathrm{T}} = \frac{D_{\mathrm{m}}}{2}\begin{bmatrix} -\dfrac{\Delta x(1-t)}{\Delta x^2 + \Delta z^2} & \dfrac{\Delta x(1-t)}{\Delta x^2 + \Delta z^2} & \dfrac{\Delta x(1+t)}{\Delta x^2 + \Delta z^2} & -\dfrac{\Delta x(1+t)}{\Delta x^2 + \Delta z^2} \\ -\dfrac{1-s}{\Delta y} & -\dfrac{1+s}{\Delta y} & \dfrac{1+s}{\Delta y} & \dfrac{1-s}{\Delta y} \\ -\dfrac{\Delta z(1-t)}{\Delta x^2 + \Delta z^2} & \dfrac{\Delta z(1-t)}{\Delta x^2 + \Delta z^2} & \dfrac{\Delta z(1+t)}{\Delta x^2 + \Delta z^2} & -\dfrac{\Delta z(1+t)}{\Delta x^2 + \Delta z^2} \end{bmatrix} \tag{A7}$$

Inserting the above results into Equation (A5) and simplifying by taking the limit $s = +1$ yields

$$\mathbf{q}_7 = \frac{\Delta y D_{\mathrm{m}}}{3\sqrt{\Delta x^2 + \Delta z^2}}\begin{bmatrix} -1 & 1 & \dfrac{1}{2} & -\dfrac{1}{2} \end{bmatrix}^{\mathrm{T}} \quad \text{and} \quad \mathbf{q}_8 = \frac{\Delta y D_{\mathrm{m}}}{3\sqrt{\Delta x^2 + \Delta z^2}}\begin{bmatrix} -\dfrac{1}{2} & \dfrac{1}{2} & 1 & -1 \end{bmatrix}^{\mathrm{T}} \tag{A8}$$

For the sake of simplicity we consider the case $\Delta z = \Delta y = \Delta x$. The element stiffness matrices are found enforcing

$$\mathbf{A}^e = \int_{-1}^{+1}\int_{-1}^{+1} [\mathbf{BDB}^{\mathrm{T}} + \mathbf{Nv}\cdot\mathbf{B}^{\mathrm{T}}]\sqrt{\det \mathbf{h}}\, ds dt = \frac{D_{\mathrm{m}}}{\sqrt{2}}\begin{bmatrix} 1-\dfrac{v\Delta x}{3D_{\mathrm{m}}} & \dfrac{v\Delta x}{3D_{\mathrm{m}}} & -\dfrac{1}{2}+\dfrac{v\Delta x}{6D_{\mathrm{m}}} & -\dfrac{1}{2}-\dfrac{v\Delta x}{6D_{\mathrm{m}}} \\ -\dfrac{v\Delta x}{3} & 1+\dfrac{v\Delta x}{3D_{\mathrm{m}}} & -\dfrac{1}{2}+\dfrac{v\Delta x}{6D_{\mathrm{m}}} & -\dfrac{1}{2}-\dfrac{v\Delta x}{6D_{\mathrm{m}}} \\ -\dfrac{1}{2}-\dfrac{v\Delta x}{6D_{\mathrm{m}}} & -\dfrac{1}{2}+\dfrac{v\Delta x}{6D_{\mathrm{m}}} & 1+\dfrac{v\Delta x}{3D_{\mathrm{m}}} & -\dfrac{v\Delta x}{3D_{\mathrm{m}}} \\ -\dfrac{1}{2}-\dfrac{v\Delta x}{6D_{\mathrm{m}}} & -\dfrac{1}{2}+\dfrac{v\Delta x}{6D_{\mathrm{m}}} & \dfrac{v\Delta x}{3D_{\mathrm{m}}} & 1-\dfrac{v\Delta x}{3D_{\mathrm{m}}} \end{bmatrix} \tag{A9}$$

After assembling of the three elements and reduction of the system accounting for the Dirichlet constraints $\psi_1 = \psi_2 = 0$, the global stiffness matrix is

$$\mathbf{A} = \frac{D_m}{\sqrt{2}} \begin{bmatrix} 2 & -1 & \frac{v\Delta x}{3D_m} & -\frac{1}{2}+\frac{v\Delta x}{6D_m} & 0 & 0 \\ -1 & 2 & -\frac{1}{2}+\frac{v\Delta x}{6D_m} & \frac{v\Delta x}{3D_m} & 0 & 0 \\ -\frac{v\Delta x}{3D_m} & -\frac{1}{2}-\frac{v\Delta x}{6D_m} & 2 & -1 & \frac{v\Delta x}{3D_m} & -\frac{1}{2}+\frac{v\Delta x}{6D_m} \\ -\frac{1}{2}-\frac{v\Delta x}{6D_m} & -\frac{v\Delta x}{3D_m} & -1 & 2 & -\frac{1}{2}+\frac{v\Delta x}{6D_m} & \frac{v\Delta x}{3D_m} \\ 0 & 0 & -\frac{v\Delta x}{3D_m} & -\frac{1}{2}-\frac{v\Delta x}{6D_m} & 1+\frac{v\Delta x}{3D_m} & -\frac{1}{2}+\frac{v\Delta x}{6D_m} \\ 0 & 0 & -\frac{1}{2}-\frac{v\Delta x}{6D_m} & -\frac{v\Delta x}{3D_m} & -\frac{1}{2}+\frac{v\Delta x}{6D_m} & 1+\frac{v\Delta x}{3D_m} \end{bmatrix} \quad \text{(A10)}$$

Taking Equation (A8) to correct the lines of **A** results in

$$\mathbf{A}_c = \frac{D_m}{\sqrt{2}} \begin{bmatrix} 2 & -1 & \frac{v\Delta x}{3D_m} & -\frac{1}{2}+\frac{v\Delta x}{6D_m} & 0 & 0 \\ -1 & 2 & -\frac{1}{2}+\frac{v\Delta x}{6D_m} & \frac{v\Delta x}{3D_m} & 0 & 0 \\ -\frac{v\Delta x}{3D_m} & -\frac{1}{2}-\frac{v\Delta x}{6D_m} & 2 & -1 & \frac{v\Delta x}{3D_m} & -\frac{1}{2}+\frac{v\Delta x}{6D_m} \\ -\frac{1}{2}-\frac{v\Delta x}{6D_m} & -\frac{v\Delta x}{3D_m} & -1 & 2 & -\frac{1}{2}+\frac{v\Delta x}{6D_m} & \frac{v\Delta x}{3D_m} \\ 0 & 0 & \frac{1}{3}-\frac{v\Delta x}{3D_m} & -\frac{1}{3}-\frac{v\Delta x}{6D_m} & \frac{2}{3}+\frac{v\Delta x}{3D_m} & -\frac{2}{3}+\frac{v\Delta x}{6D_m} \\ 0 & 0 & -\frac{1}{3}-\frac{v\Delta x}{6D_m} & \frac{1}{3}-\frac{v\Delta x}{3D_m} & -\frac{2}{3}+\frac{v\Delta x}{6D_m} & \frac{2}{3}+\frac{v\Delta x}{3D_m} \end{bmatrix} \quad \text{(A11)}$$

Inversion of the non-corrected matrix **A** and assembling of the global load vector **f** accounting for the unit source term, followed by the matrix-vector operation $\mathbf{A}^{-1}\mathbf{f}$ yields the diffusion-dependent solution

$$\boldsymbol{\psi} = \mathbf{A}^{-1} \frac{\Delta x^2}{\sqrt{2}} \begin{bmatrix} 1 \\ 1 \\ 1 \\ 1 \\ \frac{1}{2} \\ \frac{1}{2} \end{bmatrix} = \frac{\Delta x}{v} \begin{bmatrix} c_1 \\ c_1 \\ 2c_2 \\ 2c_2 \\ 3c_3 \\ 3c_3 \end{bmatrix} \quad , \quad \begin{aligned} c_1 &= \frac{v^3\Delta x^3 + 2v^2\Delta x^2 D_m + 5v\Delta x D_m^2}{(v\Delta x + D_m)^3} \\ c_2 &= \frac{v^3\Delta x^3 + 3v^2\Delta x^2 D_m + 4v\Delta x D_m^2}{(v\Delta x + D_m)^3} \\ c_3 &= \frac{v^3\Delta x^3 + 8v^2\Delta x^2 D_m/3 + 3v\Delta x D_m^2}{(v\Delta x + D_m)^3} \end{aligned} \quad \text{(A12)}$$

which is obviously not correct, while inversion of the corrected matrix $\mathbf{A}_c$ produces the diffusion-independent correct solution

$$\boldsymbol{\psi}_c = \mathbf{A}_c^{-1}\mathbf{f} = \frac{\Delta x}{v}\begin{bmatrix} 1 & 1 & 2 & 2 & 3 & 3 \end{bmatrix}^{\mathrm{T}} \quad \text{(A13)}$$

In Figure A2 the difference between the two schemes is illustrated. The error induced by the homogeneous Neumann condition at outlet naturally increases with dispersion.

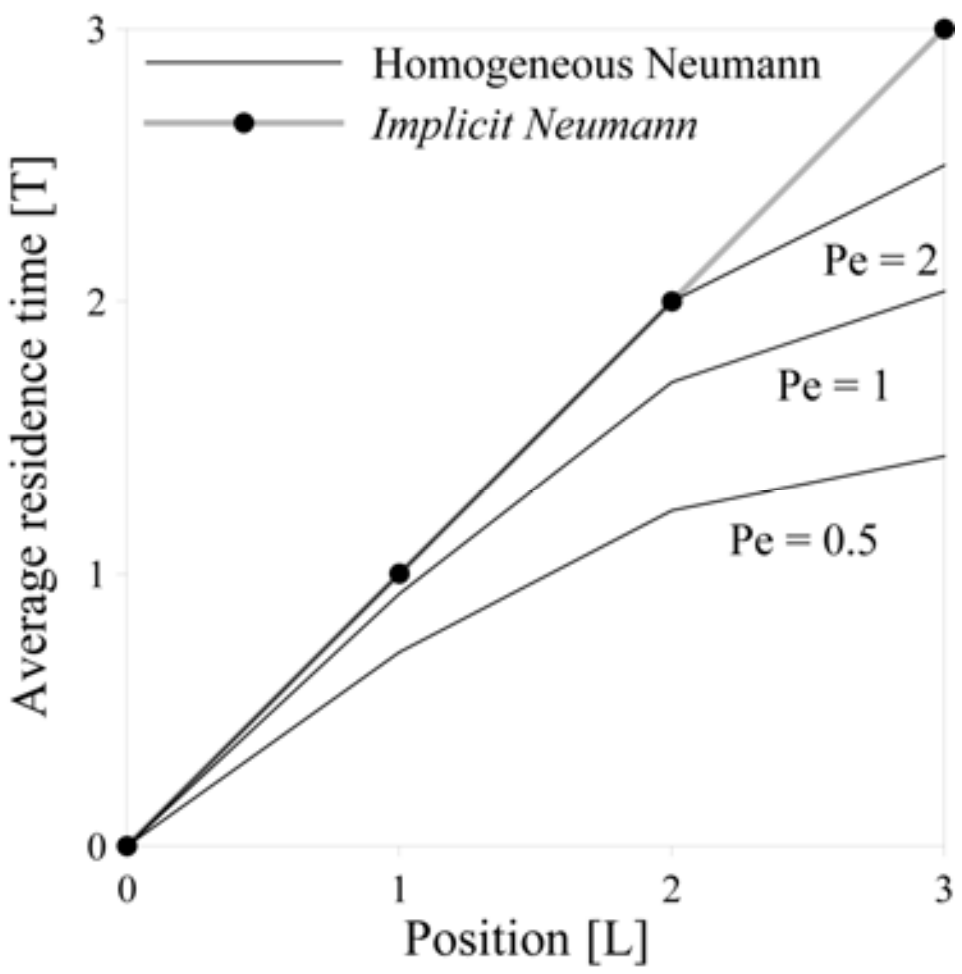

Figure A2. Convective-diffusive transport of the average residence time within a fracture in the 3-D space. Solution for three Péclet numbers $Pe = v\Delta r/D_m$ (indicated on the figure), for the case $\Delta x = \Delta y = \Delta z = 1/\sqrt{2}$, $\mathbf{v} = [1/\sqrt{2} \quad 0 \quad 1/\sqrt{2}]$, $\Delta r = \sqrt{(\Delta x^2 + \Delta z^2)}$, $v = \|\mathbf{v}\|$.